\documentclass[10pt,aps,pre,onecolumn,notitlepage,superscriptaddress,showpacs,preprintnumbers]{revtex4-1}

\usepackage{amsmath,amssymb,amsfonts}
\usepackage{graphicx,color}
\usepackage{verbatim}
\usepackage{float}
\usepackage{dcolumn}
\usepackage{natbib}
\usepackage{bm}
\usepackage{hyperref}
\usepackage{multirow}

\newcommand{\ie}{\emph{i.e.}}
\newcommand{\eg}{\emph{e.g.}}

\newcommand{\avg}[1]{\langle #1\rangle}

\newcommand{\be}{\begin{equation}}
\newcommand{\ee}{\end{equation}}
\newcommand{\bea}{\begin{eqnarray}}
\newcommand{\eea}{\end{eqnarray}}

\begin{document}

\title{Influence of technological innovations on industrial production:\\A motif analysis on the multilayer network}

\author{Martina Formichini}
\affiliation{Dipartimento di Fisica, ``Sapienza'' Università di Roma, Piazzale A. Moro 2 - 00185 Rome (Italy)}
\author{Giulio Cimini}
\affiliation{IMT School for Advanced Studies, Piazza S.Francesco 19, 55100 Lucca (Italy)}
\affiliation{Istituto dei Sistemi Complessi (ISC) - CNR UoS Sapienza, Dipartimento di Fisica, ``Sapienza'' Università di Roma, Piazzale A. Moro 2 - 00185 Rome (Italy)}
\author{Emanuele Pugliese}
\affiliation{EC - JRC, Calle Inca Garcilaso 3 - E-41092 Seville (Spain)}
\thanks{The content of this article does not reflect the official opinion of the European Union. Responsibility for the information and views expressed therein lies entirely with the authors.}
\author{Andrea Gabrielli}
\affiliation{Istituto dei Sistemi Complessi (ISC) - CNR UoS Sapienza, Dipartimento di Fisica, ``Sapienza'' Università di Roma, Piazzale A. Moro 2 - 00185 Rome (Italy)}
\affiliation{IMT School for Advanced Studies, Piazza S.Francesco 19, 55100 Lucca (Italy)}

\begin{abstract}
In this work we study whether specific combinations of technological advancements can signal the presence of local capabilities allowing for a given industrial production. 
To this end, we generate a multi-layer network using country-level patent and trade data, and perform a motif-based analysis on this network using a statistical validation approach derived from maximum entropy arguments. 
We show that in many cases the signal far exceeds the noise, providing robust evidence of synergies between different technologies that can lead to a competitive advantage in specific markets. 
Our results can be highly useful for policy makers to inform industrial and innovation policies.
\end{abstract}

\maketitle

\section{Introduction}

Technological innovation is the main driver of modern economic growth \cite{freeman1989technology,romer1990endogeneous}.
It is therefore not surprising that measuring and predicting the potential impact of technological innovations on export competitiveness has been the central issue of many studies in the last forty years \cite{krugman1979model,soete1980impact,dosi1990,verspagen1992} as well as the focus of a general interest in the field of innovation systems \cite{nelson1993national}. Several empirical studies that tried to measure such effects of national innovativeness on productivity and trade were also carried out, with mixed results \cite{griffith2004mapping,bronwyn2010}.
Overall, such academic effort provided a theoretical framework and empirical stylized facts helpful to understand the aggregate effect of innovation in determining the competitive advantage of countries in different markets. 
Policy makers are however more interested in identifying the specific technologies that are relevant for specific markets \cite{junker2015,EC2009}---a task that is much harder to deal with in an organic and objective fashion. Indeed, the scientific effort addressing the impact of specific technologies on specific markets has been limited to {\em ad hoc} case studies that are difficult to compare \cite{Devaraj2000, Sher2005}.

A recent paper of ours \cite{multilayer} deals with this issue using a multilayer network characterization of the innovation system. Specifically, a three-layered network of innovation activities is derived starting from the three bipartite networks describing the scientific, technological and production activities of countries; the connection of the multilayer network represent the conditional probability that the information produced by an innovation activity (\eg, a technological sector) will be used in another innovation activity (\eg, an industrial product category) after a given time. Grounded on previous fundamental studies of Economic Complexity \cite{EFC1, EFC2}, this is the first attempt to build a representation of the innovation system as a complex multilayer network.

In this paper we generalize this approach by measuring the potential influence that a pair of activities has on another activity. That is, we go beyond the single link analysis and consider the {\em motifs} of the multilayer network \cite{Milo2002,Saracco2015}. For simplicity we limit our analysis to the relationships between (pairs of) technologies and products, which however represent a crucial aspect of the innovation system---since the interaction between different technologies is often a driver of innovation and progress \cite{recombinant}. As in \cite{multilayer}, we carry out a statistical validation of our results against a null network models derived from maximum entropy principles.

\section{Materials and methods}\label{MM}

In order to build the bi-layered network of technologies and products that will be used in our analysis, we start from the following popular databases.

PATSTAT (\url{www.epo.org/searching-for-patents/business/patstat}) collects all the patents by different Patent Offices around the world. The basic units of observation in the dataset is a patent family (\ie, the set of patents with common priorities, that is, referred to the same innovation). Each family is related to the countries of origin of the applicants, and to a (set of) technological codes defined by the International Patent Classification (IPC). We define $W_{ct}(y)$ as the number of patent families associated to IPC code $t$ applied by firms located in country $c$ on year~$y$.

BACI export data, recorded by UN COMTRADE (\url{https://comtrade.un.org/}), collects the import-export flows (quantified in thousands of current US dollars) among countries in the world, related to production as classified using the Harmonized System 2007 of the World Customs Organization. 
We define $W_{cp}(y)$ as the monetary value of the overall export of country $c$ for product $p$ during year $y$. Note that we use export data as proxies of (competitive) industrial production, as typically done in the Economic Complexity literature.

In this work we consider a time span of data ranging from 1995 to 2012, for which we have a reliable coverage for both patent and export data. For technologies, we use a 4-digits resolution of IPC codes, resulting in a number of technological sectors $N_t$ ranging between 629 and 636. For products, we again use a 4-digit resolution of the Harmonized System, resulting in a number of product categories $N_p$ ranging between 1140 and 1176. Finally, the number of considered countries $N_c$ varies between 66 and 72. The slight variations of these numbers depend on the particular year considered, and are due to geopolitical changes and periodical re-categorization of technologies and products.

Using these basic data, we can define the Revealed Comparative Advantage (RCA) \cite{RCA1965} of a country $c$ on an activity $a$ (which is either a technological sector $t$ or and product category $p$) in a given year $y$:
\be
RCA_{ca}=\dfrac{W_{ca}(y)}{\sum_{a'=1}^{N_a}W_{ca'}(y)}\Bigg/\dfrac{\sum_{c'=1}^{N_c}W_{c'a}(y)}{\sum_{c'=1}^{N_c}\sum_{a'=1}^{N_a}W_{c'a'}(y)}\,.
\ee
Thanks to the RCA we can further define for each year $y$ the binary bipartite networks countries-technologies and countries-products. These are respectively represented by the binary biadjacency matrices $\mathbf{M}^{\mathcal{C,T}}(y)$ and $\mathbf{M}^{\mathcal{C,P}}(y)$ whose elements $M_{ct}(y)$ and $M_{cp}(y)$ are:
\be
M_{ca}(y)=\left\{
\begin{array}{ll}
1 & \mbox{if }RCA_{ca}(y)\ge 1\\
& \\
0 &\mbox{otherwise}
\end{array}
\right.
\ee
(where again $a$ refers to a technological sector $t$ in the case of the bipartite countries-technologies network and to a product category $p$ in the countries-product case). 

Once we have the matrices $\mathbf{M}^{\mathcal{C,T}}(y_1)$ for the year $y_1$ and $\mathbf{M}^{\mathcal{C,P}}(y_2)$ for the year $y_2$, in analogy with \cite{multilayer}, we can construct the {\em assist matrix} $\mathbf{B}^{\mathcal{T\to P}}(y_1,y_2)$, whose generic element is defined as:
\begin{equation}
B_{tp}(y_1,y_2)=\sum_{c=1}^{N_c}\frac{M_{ct}(y_1)}{k_t(y_1)}\frac{M_{cp}(y_2)}{k_c^{(p)}(y_2)}\,.
\label{assist}
\end{equation}
In the above expression, $k_t(y_1)=\sum_{c=1}^{N_c} M_{ct}(y_1)$ is the number of countries having technology $t$ in their technological portfolio at year $y_1$, and $k_c^{(p)}(y_2)=\sum_{p=1}^{N_p} M_{cp}(y_2)$ is the cardinality of the product basket of country $c$ in year $y_2$.
As explained in \cite{multilayer}, $B_{tp}(y_1,y_2)$ with $y_1\le y_2$ gives the conditional probability that a bit of information produced in the technological sector $t$ in year $y_1$ arrives (via a random walk on the coupled bipartite network) at the product category $p$ in year $y_2$, through one of the countries having $t$ in its technological basket at $y_1$ and $p$ in its product basket at $y_2$. The elements $B_{tp}(y_1,y_2)$ then represent the weighted links of the bi-layered (or bipartite) network technologies-products, with the former at year $y_1$ and the latter at year $y_2$. These links are extensively studied in \cite{multilayer}.

Here we move forward and consider the $\Lambda$ motifs of the bipartite technologies-products network:
\be
\Lambda_{tt'}^{p}(y_1,y_2)=B_{tp}(y_1,y_2)B_{t'p}(y_1,y_2).
\label{Lambda}
\ee 
$\Lambda_{tt'}^{p}(y_1,y_2)$ gives the conditional probability that two bits of information originally located on technologies $t$ and $t'$ respectively at year $y_1$ {\em both} reach product $p$ at year $y_2$. In other words, this motif quantifies the joint probability for the co-occurrence in a single country of the pair technology $t$ and product $p$ {\em and} of the pair technology $t'$ and product $p$, where the two events are considered as independent. 
Note that while this interpretation of the $\Lambda$ motifs cannot be directly related to ``impact'' or ``causality'', it does go beyond a simpler measure of a (time-dependent) correlation. 
Note also that the name $\Lambda$ motif comes from the fact that, in the bipartite network technologies-products, this quantity gives the weight of the a $\Lambda$ shaped set of two links having different origin in the technology layer and same end in the product layer \cite{Saracco2015}. In principle it is possible to generalize this approach by considering higher-order motifs, \eg, by assessing the influence of a wider group of technologies on a single product. For the sake of simplicity and limits of statistical significance, here we focus on the simple $\Lambda$ motif. 

After obtaining the empirical values of the $\Lambda$ motifs from data, we statistically validate them using their probability distribution derived from an appropriate null model, which is schematically defined as follows (see Appendix for a thorough presentation). 
For each year $y$, we build two statistical ensembles of biadjacency matrices $\mathbf{\tilde M}^{\mathcal{C,T}}(y)$ and  $\mathbf{\tilde M}^{\mathcal{C,P}}(y)$ respectively for the  bipartite networks countries-technologies and countries-products. These networks are built such to be maximally random, apart from having the ensemble average of node degrees equal to the values observed in the empirical networks. For node degrees, we mean both the technological diversification of countries $\tilde k_c^{(t)}(y)=\sum_{t}\tilde M_{ct}^{y}$ and the technologies ubiquities $\tilde k_t(y)=\sum_{c}\tilde M_{ct}^{y}$ for the countries-technologies network, and both the product diversification of countries $\tilde k_c^{(p)}(y)=\sum_{p}\tilde M_{cp}^{y}$ and the product ubiquities $\tilde k_p(y)=\sum_{c}\tilde M_{cp}^{y}$ for the countries-products network. 
We choose these quantities as constraints as we want our null model to bear only the information contained in the diversification of countries and the ubiquity of activities, without taking into account the specific pattern of co-occurrences found in the empirical networks. 
In the spirit of the information theory interpretation of statistical mechanics \cite{Jaynes,Park2004,Cimini2018}, the probability measure defining both statistical ensembles of binary bipartite networks is obtained using a constrained entropy maximization approach. The resulting ensembles are known in the literature as Bipartite Configuration Models (BiCM) \cite{Saracco2015}. 
Finally, given the BiCM ensembles for the bipartite networks $\mathbf{\tilde M}^{\mathcal{C,T}}(y_1)$ and  $\mathbf{\tilde M}^{\mathcal{C,P}}(y_2)$, we can use Eqs.~\eqref{assist} and \eqref{Lambda} appropriately applied to BiCM quantities to derive the probability distribution for the value $\tilde\Lambda_{tt'}^p(y_1,y_2)$ in the null model \cite{Gualdi2016,Saracco2017}. Numerically, we populate the BiCM ensembles by generating $10^3$ matrices $\tilde{\mathbf{M}}^{\mathcal{C,T}}(y_1)$ and $\tilde{\mathbf{M}}^{\mathcal{C,P}}(y_2)$, and then contract each pair to generate a final ensemble of $10^3$ null matrices $\tilde{\mathbf{B}}^{\mathcal{T\to P}}(y_1,y_2)$. 

In the following we will focus on $\Lambda_{tt'}^{p}(\Delta y)$, that is, the average value of $\Lambda_{tt'}^{p}(y_1,y_2)$ over all pairs of years giving the same difference $y_2-y_1=\Delta y$. 
This represents the conditional probability that two bits of information, produced in the same year for a pair of technologies $t$, $t'$, reach a product $p$ after $\Delta y$ years. 
We define the signal $\phi(\Delta y)$ as the fraction of significant $\Lambda_{tt'}^{p}(\Delta y)$ (at the $\alpha=0.01$ significance level, according to the probability distribution of the null model) for combinations of $t$, $t'$ and $p$ chosen for selected matrix regions. In general we consider a population of motifs equal to 5500 units (see below). 

\section{Results}\label{Res}

\begin{figure}[h!]
\centering
\includegraphics[width=0.5\textwidth]{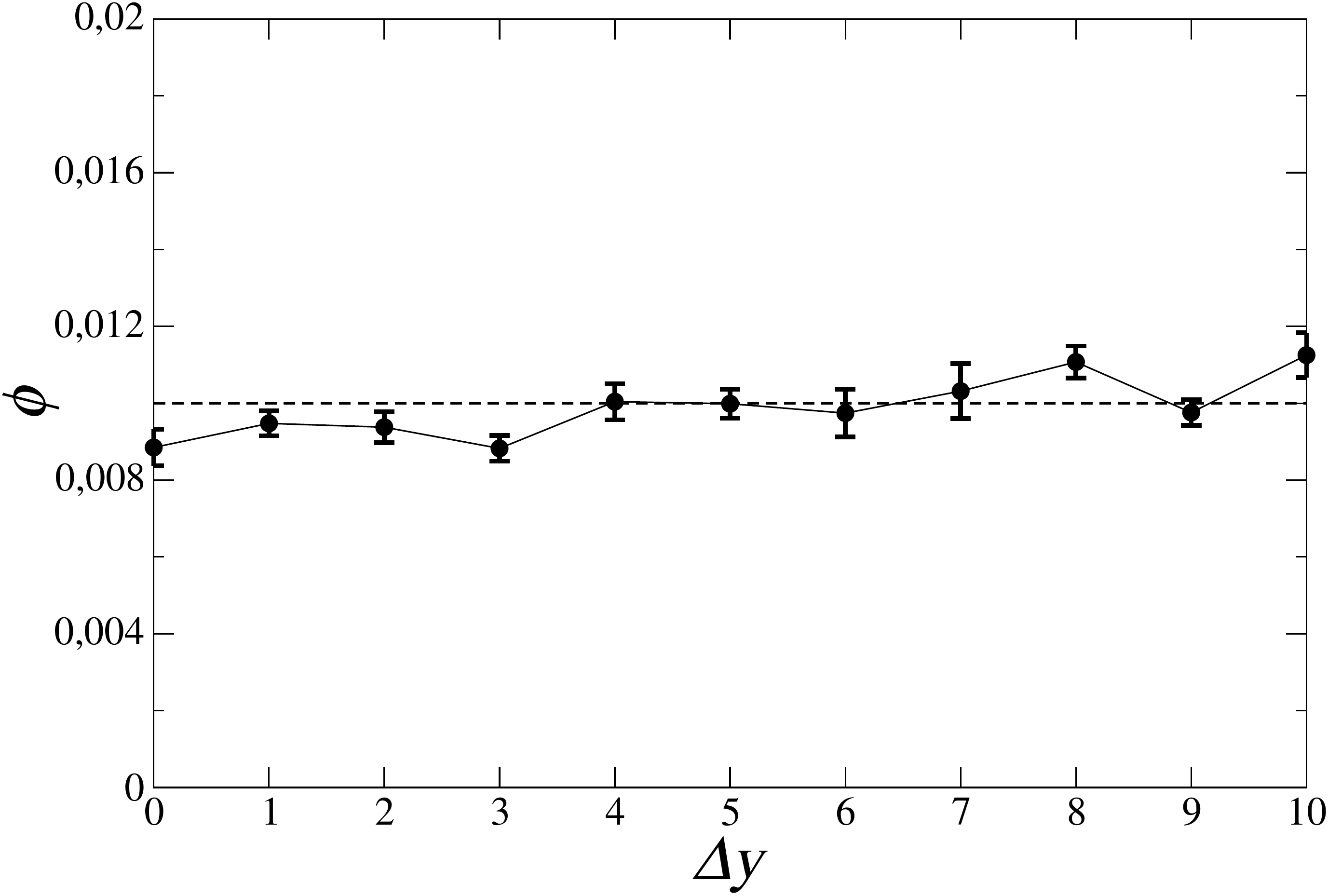}
\caption{Mean signal $\phi$ computed over 5500 combinations of $t$, $t'$ and $p$ chosen at random, for different values of the time lag $\Delta y$. 
Error bars represent the standard deviation over the year pairs giving the same time lag, whereas, the dotted line is the significance level $\alpha$.}
\label{fig1}
\end{figure}

As first test we report the mean signal $\phi$, that is, the signal averaged over all combinations of $t$, $t'$ and $p$. 
Since the total number of such motifs is extremely large, we choose 5500 motifs at random and take their mean signal as representative of the global average. 
Figure \ref{fig1} shows that $\phi$ basically remains within one standard deviation from the noise level $\alpha$, indicating that the mean signal within the data is negligible.

\begin{figure}[h!]
\centering
\includegraphics[width=0.5\textwidth]{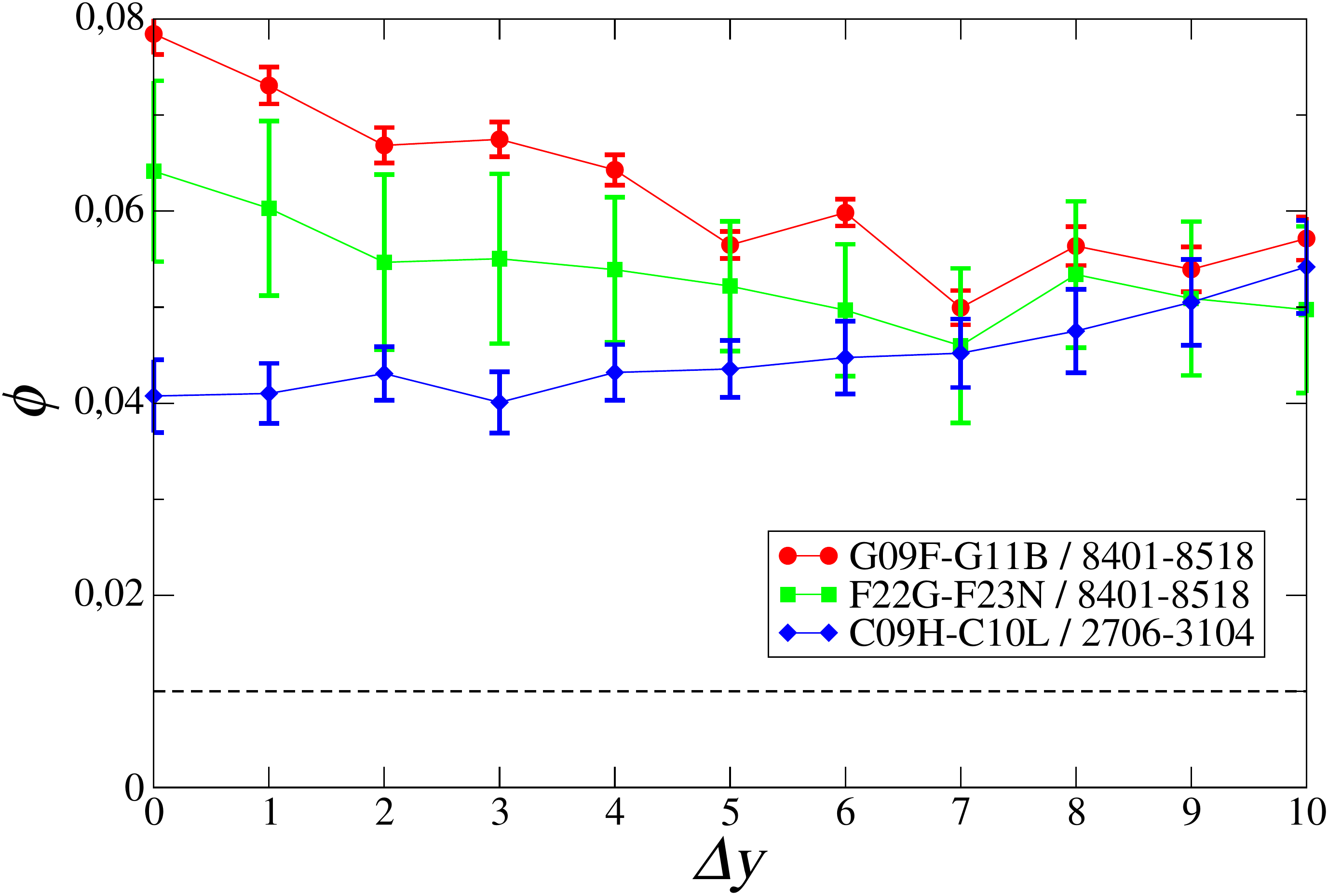}
\caption{Mean signal $\phi$ computed over 5500 combinations of $t$, $t'$ and $p$ chosen for specific region of the assist matrix. 
Red circles: technological codes related to sector ``physics'' -- in the region G09F-G11B {\em instruments of communications, acoustics, optics}, 
and products in the region 8401-8518 {\em machinery and metals}. 
Green squares: technological codes related to sector ``engineering'' -- in the region F22G-F23N {\em various types of machines including steam and combustion}, and products in the region 8401-8518 {\em machinery and metals}. 
Blue diamonds: technological codes related to sector ``chemistry'' -- C09H e C10L {\em macromolecular and inorganic compounds, gas and petroleum}, products, and products in the region 2706-3104 {\em inorganic and organic chemicals, pharmaceuticals}. 
In all cases, error bars represent the standard deviation over the year pairs giving the same time lag, whereas, the dotted line is the significance level $\alpha$.}
\label{figX}
\end{figure}

We then report the signal relative to motifs within selected regions of the assist matrix (Fig.~\ref{figX}). Specifically, we choose sub-regions 
of 11 technologies and 100 products (related to specific technological sectors and product categories), whose total number of $\Lambda$ motifs is 5500 (since $\Lambda_{tt'}^p=\Lambda_{t't}^p$). 
From Fig.~\ref{figX} we see that, by selecting coherent sets of technologies and products, the signal is well enhanced: 
the presence of a pair of technologies in the capability basket of a country can predict whether that country can successfully export a product, 
and this happens almost independently on the time lag $\Delta y$. 

\begin{figure}[h!]
\centering
\includegraphics[width=0.5\textwidth]{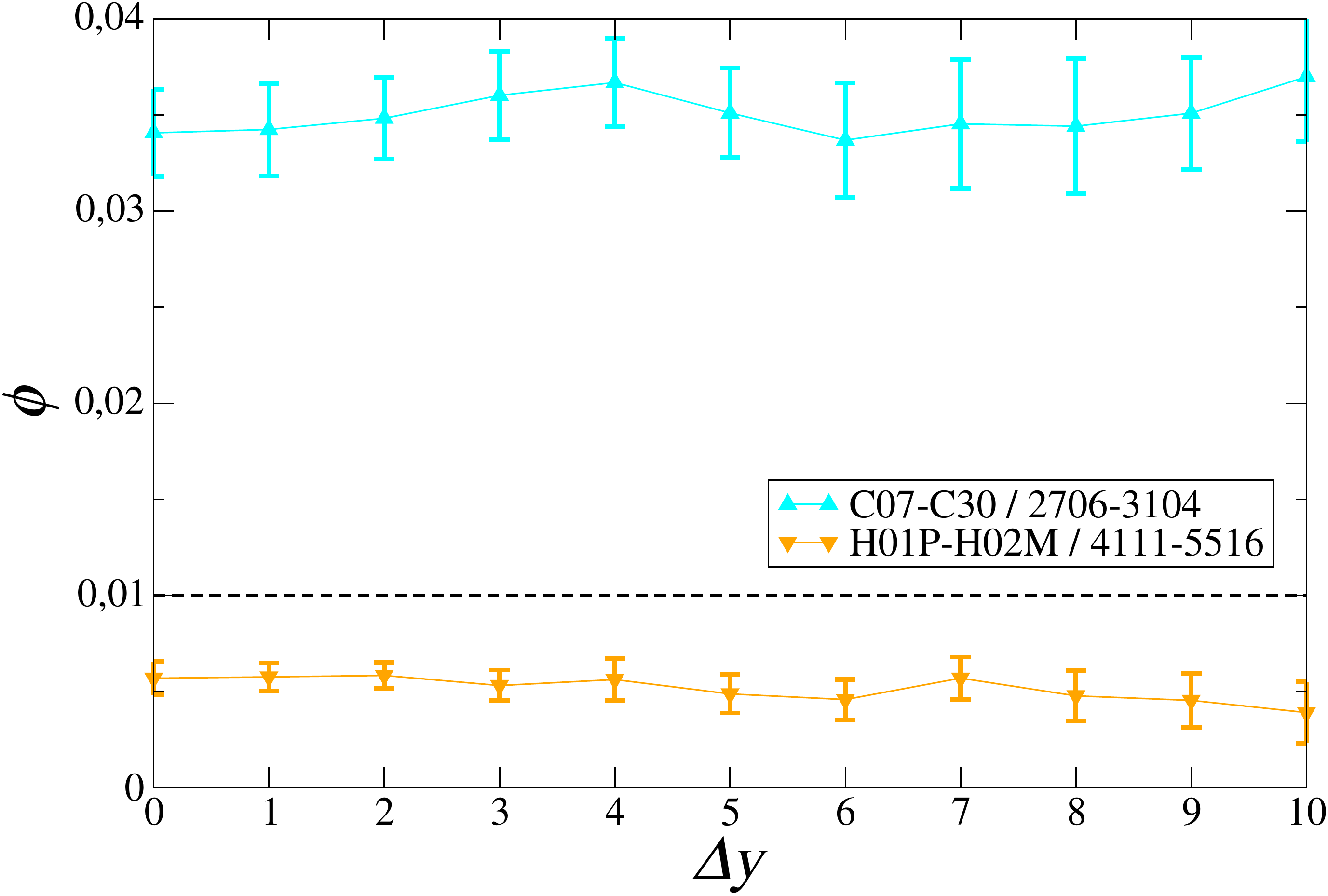}
\caption{Mean signal $\phi$ computed over 5500 combinations of $t$, $t'$ and $p$ chosen for specific region of the assist matrix. 
Cyan upper triangles: technological codes related to sector ``chemistry'' -- C07 e C30 (\ie, a much larger region than that represented in the bottom panel of Figure \ref{figX}) and products in the region 2706-3104 {\em inorganic and organic chemicals, pharmaceuticals}. 
Orange lower triangles: technological codes related to sector ``physics'' -- in the region H01P-H02M {\em electricity}, and products in the region 4411-5516 {\em textiles}.
In all cases, error bars represent the standard deviation over the year pairs giving the same time lag, whereas, the dotted line is the significance level $\alpha$.}
\label{figY}
\end{figure}

As consistency checks we make two exercises, both reported in Figure \ref{figY}. Firstly we show that the results we just presented 
do not depend on the particular resolution used to choose the motifs. Secondly we show that for incoherent technologies and products we indeed get a much lower signal---even lower than the significance level. In this latter case, a significant development of specific technologies 
corresponds to a low level of export for given products.

We finally provide a few examples of motifs with high signal (\ie, with low p-value). To do that, since the total number of motifs is extremely high 
so that a complete exploration cannot be performed efficient, we considered the motifs made up of the link pairs $B_{tp}(y_1,y_2)$ and $B_{t'p}(y_1,y_2)$ 
which are independently those with the highest signal. Table \ref{tab1} reports some instances of such motifs for the specific choice $\Delta y=0$. 
The triplets $t,t',p$ appearing in the Table seem indeed coherent, and confirm that our method can actually extract meaningful information in an unsupervised way.

\begin{table}
\begin{tabular}{l|l}
p-value & $p$, $t$, $t'$\\
\hline\hline
\multirow{3}{*}{$2\cdot 10^{-4}$} & 4701: Wood pulp\\
& C05B: Lime; Magnesia; Slag; Cements\\
& C09K: Materials for applications not otherwise provided for\\
\hline
\multirow{3}{*}{$5\cdot 10^{-4}$} & 2605: Mineral products\\
& C21D: Modifying the physical structure of ferrous metals\\
& F04F: Pumping of fluid by direct contact of another fluid or by using inertia of fluid to be pumped\\ 
\hline
\multirow{3}{*}{$5\cdot 10^{-4}$} & 2605: Mineral products\\
& C21D: Modifying the physical structure of ferrous metals\\
& F04F: Working metallic powder\\
\hline
\multirow{3}{*}{$5\cdot 10^{-4}$} & 8443: Printing machine\\
& D02H: Mechanical methods or apparatus in the manufacture of artificial filaments\\
& G01T Measurement of nuclear or x-radiation\\
\hline
\multirow{3}{*}{$7\cdot 10^{-4}$} & 4703: Chemical wood pulp\\
& D21F: Decorating textiles\\
& B27C: Planing, drilling, milling, turning, or universal machines \\
\hline
\multirow{3}{*}{$8\cdot 10^{-4}$} & 2605: Mineral products\\
& C21D: Modifying the physical structure of ferrous metals\\
& FF15B: Systems acting by means of fluids in general\\
\hline
\multirow{3}{*}{$2\cdot 10^{-3}$} & 4703: Chemical wood pulp\\
& D21F: Paper-making machines\\
& F03D: Wind motors\\
\hline
\multirow{3}{*}{$2\cdot 10^{-3}$} & 4703: Chemical wood pulp\\
& D21F: Paper-making machines\\
& D06Q: Decorating textiles\\
\hline
\multirow{3}{*}{$3\cdot 10^{-3}$} & 8519: Sound recording or reproducing apparatus\\
& G10K: Sound-producing devices\\
& G01T: Capacitors; rectifiers, detectors, switching devices\\
\hline
\multirow{3}{*}{$3\cdot 10^{-3}$} & 8519: Sound recording or reproducing apparatus\\
& G10K: Sound-producing devices\\
& G04f: Time-interval measuring\\
\hline\hline
\end{tabular}
\caption{Examples of highly significant $\Lambda$ motifs. The p-value is averaged over all year pairs $y_1=y_2$ giving $\Delta y=0$. 
To make this selection, we picked the most significant pairs (individual links) $(t,p)$ 
and then choose $t'$ within the region where the average p-value was highest.}
\label{tab1}
\end{table}

\section{Conclusions}

In this work we provide an effectual way of measuring the combined effect of a set of technologies on one product. In particular we consider lambda motifs, quantifying the paired effect of two technology together. In the process of finding relevant combinations, we highlight several results. 

First of all, we show how the combination of multiple technologies has a very different role in different industrial and technological sectors. While technologies related to physics, engineering and chemistry tend to show synergies between them in enhancing the chances of successful export of a product (Figure 2), looking at two generic technologies there is no such effect (Figure 1). This heterogeneity, while expected, is here quantitatively measured. Secondly, we confirm that co-occurrences between technological activities in a country are able to extract information on shared capabilities, to inform policy makers and stakeholders of relevant synergies, like those highlighted in Table 1, for specific export markets.

The mapping provided by our approach for the effects of pairs of technologies on products can represent the fundamental building block for the formulation of a powerful instrument to inform policies and industrial strategies about technology transfer. This operational step will be an important aspect of future research.

\section{Appendix: the Bipartite Configuration Model (BiCM) and the null model of the Assist matrix}
\label{BiCM-L}

In order to assess the statistical significance of elements of the assist matrices, we resort to a null model for the bipartite matrices $\mathbf{M}^{\cal C, T}(y)$ and $\mathbf{M}^{\cal C, P}(y)$, 
built by randomly reshuffling their elements (\ie, the network links connecting nodes in the layer ${\cal C}$ of countries to nodes in the layers ${\cal T}$ of technologies and ${\cal P}$ of products, respectively), but preserving the diversification of countries and the ubiquity of the different innovation activities (\ie, the degrees of the nodes in both layers of the bipartite networks countries-technologies and countries-products) . 
This means that we randomize the signal coming from the network connectivity patters beyond that encoded in the nodes degrees. 
In order to analytically formulate the null model, avoiding to rely on a conditional uniform graph test~\cite{Zweig2011,Neal2014}, degree constraints are imposed on average, in a way formally similar to what happens for the canonical ensemble in Statistical Mechanics with the constraint on the energy \cite{Jaynes}. 
This amounts to set a null hypothesis described by the {\em Bipartite Configuration Model} (BiCM)~\cite{Saracco2015}, which is
an extension of the {\em Configuration Model} \cite{Park2004} to bipartite networks.

In the following, we use symbols with the tilde for quantities assessed on null model configurations, and without the tilde for empirically observed values. 
From an operational viewpoint, the BiCM null model for any binary biadjacency matrix $\mathbf{M}$, representing a (real) {\em empirical} bipartite network with two layers of nodes $a\in A$ and $b\in B$, is built using two main steps:
\begin{enumerate}
\item Through a constrained {\em maximum entropy} approach, we define the ensemble $\Omega$ of bipartite networks which are maximally random, apart from the ensemble average of the node degrees on both layers of the bipartite network which are constrained to generic fixed values. Such an ensemble is thus an instance of Exponential Random Binary Graph (ERBG). 
\item In order to determine the ERBG that best represents the empirical bipartite network, we use a {\em maximal likelihood} argument showing that the mean values of the node degrees have to be taken equal to the observed ones in the empirical network \cite{Squartini2018}: 
$\avg{\tilde{k}_a}_{\Omega}=k_a$ $\forall a\in A$ and $\avg{\tilde{k}_b}_{\Omega}=k_b$ $\forall b\in B$, where we have indicated with $k$ the observed degrees in the real network and with $\tilde k$ the degrees in a generic configuration of the null model.  We remind that $k_{a}=\sum_b M_{ab}$ and $k_b=\sum_a M_{ab}$ and analogously for "tilded" quantities.
\end{enumerate}

Let us start by introducing the ERBG with fixed mean node degrees and let $\tilde{\mathbf{M}}\in\Omega$ be a network configuration in such ensemble and $P(\tilde{\mathbf{M}})$ be its occurrence probability. 
By implementing the prescriptions from Information Theory and  Statistical Mechanics \cite{Huang1987, Jaynes}, the least biased choice of $P(\tilde{\mathbf{M}})$ is the one that maximizes the informational entropy
\begin{equation}
 S=-\sum_{\tilde{\mathbf{M}}\in\Omega} P(\tilde{\mathbf{M}})\, \ln P(\tilde{\mathbf{M}}), \label{eq:entropy}
\end{equation}
subject to the normalization condition $\sum_{\tilde{\mathbf{M}}\in\Omega} P(\tilde{\mathbf{M}}) = 1$ plus the constraints:
\begin{equation}
 \avg{\tilde{k}_a}_{\Omega} = \sum_{\tilde{\mathbf{M}}\in\Omega} P(\tilde{\mathbf{M}})\,\tilde{k}_a(\tilde{\mathbf{M}}) = k^*_a\quad\forall a\in A,\qquad 
 \avg{\tilde{k}_b}_{\Omega} = \sum_{\tilde{\mathbf{M}}\in\Omega} P(\tilde{\mathbf{M}})\,\tilde{k}_b(\tilde{\mathbf{M}}) = k^*_b\quad\forall b\in B. \label{eq:constraints}
\end{equation}
where $k^*_a$ $\forall a \in A$ and $k^*_b$ $\forall b \in B$ are arbitrarily fixed values for the mean degrees of nodes belonging to layers $A$ and $B$.
By defining the respective Lagrange multipliers $\omega$, $\{\mu_a\}_{a\in{A}}$ and $\{\nu_b\}_{b\in{B}}$ (one for each node of the bipartite network), the probability distribution of all configurations $\tilde{\mathbf{M}}\in\Omega$ that maximizes the entropy satisfying at the same time all the constraints, is determined by the following variational equation:
\begin{equation}
\begin{split}
0\:=\;& \frac{\delta}{\delta P(\tilde{\mathbf{M}})}\left[S+\omega\left(1-\sum_{\tilde{\mathbf{M}}\in\Omega} P(\tilde{\mathbf{M}})\right)+\right.  \\
& \left.+\sum_{a\in A}\,\mu_a\left(k^*_a-\sum_{\tilde{\mathbf{M}}\in\Omega}P(\tilde{\mathbf{M}})\,\tilde{k}_a(\tilde{\mathbf{M}})\right)
+\sum_{b\in B}\,\nu_b\left(k^*_b-\sum_{\tilde{\mathbf{M}}\in\Omega}P(\tilde{\mathbf{M}})\,\tilde{k}_b(\tilde{\mathbf{M}})\right)\right].
\end{split}
\end{equation}
It is a matter of simple algebra to show that the solution of this equation is:
\begin{equation}
P(\tilde{\mathbf{M}}\,|\,\{\mu_a\},\{\nu_b\})=e^{-H(\tilde{\mathbf{M}}\,|\,\{\mu_a\},\{\nu_b\})}\Big/Z(\{\mu_a\},\{\nu_b\})\,, 
\label{eq:p_good}
\end{equation}
where the function $H(\tilde{\mathbf{M}}\,|\,\{\mu_a\},\{\nu_b\})$ is usually called the Hamiltonian of the graph configurations
\begin{equation}
H(\tilde{\mathbf{M}}\,|\,\{\mu_a\},\{\nu_b\})=\sum_{a\in A}\mu_a\,\tilde{k}_a(\tilde{\mathbf{M}})+\sum_{b\in B}\nu_b\,\tilde{k}_b(\tilde{\mathbf{M}}),\label{eq:acca}
\end{equation}
and $Z(\{\mu_a\},\{\nu_b\})$ is the corresponding partition function 
\begin{equation}
Z(\{\mu_a\},\{\nu_b\})=e^{\omega+1}=\sum_{\tilde{\mathbf{M}}\in\Omega} e^{-H(\tilde{\mathbf{M}}\,|\,\{\mu_a\},\{\nu_b\})}. \label{eq:zeta}
\end{equation}
The above Eqs. \eqref{eq:p_good}, \eqref{eq:acca} and \eqref{eq:zeta} define the network ensemble known as the BiCM model. 

Note that, as we have implemented only local constraints, \ie, the mean node degrees, Eq. \eqref{eq:p_good} can be rewritten as the product of single link probability distributions over all pair of nodes belonging respectively to the two different layers \cite{Saracco2015}:
\begin{equation}
P(\tilde{\mathbf{M}}\,|\,\{\mu_a\},\{\nu_b\})=\prod_{a\in A}\prod_{b\in B} \pi_{ab}^{\tilde{M}_{ab}}\,(1-\pi_{ab})^{\tilde{M}_{ab}}, \label{eq:p_expl}
\end{equation}
where $\pi_{ab}$ is simply the probability of the link between nodes $a\in A$ and $b\in B$:
\begin{equation}
 \pi_{ab}=\avg{\tilde{M}_{ab}}_{\Omega}=\sum_{\tilde{\mathbf{M}}\in\Omega} \tilde{M}_{ab}\,P(\tilde{\mathbf{M}}\,|\,\{\mu_a\},\{\nu_b\})=\frac{\eta_a\,\theta_b}{1+\eta_a\,\theta_b} 
 \label{conn_prob}
\end{equation}
with $\eta_a=e^{-\mu_a}$ and $\theta_b=e^{-\nu_b}$. In other words the existence of different links are independent events with respective probabilities which are function only of the Lagrange multipliers associated to the node pairs defining the links.
The values of the Lagrange multipliers are determined by the constraints Eqs.~\eqref{eq:constraints} which can be rewritten in terms of the derivatives of the partition function:
\begin{equation}
-\frac{\partial}{\partial \mu_a} \ln Z(\{\mu_a\},\{\nu_b\})\equiv\avg{\tilde{k}_a}_{\Omega}=k^*_a\quad\forall a\in A,
\qquad -\frac{\partial}{\partial \nu_b} \ln Z(\{\mu_a\},\{\nu_b\})\equiv\avg{\tilde{k}_b}_{\Omega}=k^*_a\quad\forall b\in B.
\end{equation}
Equations \eqref{eq:p_good}-\eqref{conn_prob} define the generic EBRG with fixed mean degrees of nodes on both layers of the bipartite network.

Once the generic EBRG has been defined, we can move to the step of determining, among all possible EBRGs, the optimal null model for a given {\em real} biadjacency matrix (\ie, a bipartite network) $\mathbf{M}$. Equivalently we have to choose the best values for $\{k^*_a\}_{a\in A}$ and $\{k^*_b\}_{b\in B}$, \ie, for the Lagrange multipliers $\{\mu_a\}_{a\in A}$ and $\{\nu_b\}_{b\in B}$, in relation to the connectivity properties of $\mathbf{M}$.

To this aim we write the log-likelihood function \cite{Squartini2018}
\begin{equation}
\mathcal{L}(\{\mu_a\},\{\nu_b\})=\ln P(\mathbf{M}\,|\,\{\mu_a\},\{\nu_b\})=
\sum_{a\in A}k_a\,\ln\eta_a+\sum_{b\in B}k_b\,\ln\theta_b-\sum_{a\in A}\sum_{b\in B}\ln(1+\eta_a\,\theta_b), 
\label{eq:likelihood}
\end{equation}
where $P(\mathbf{M}\,|\,\{\mu_c\},\{\nu_a\})$ is the probability measure \eqref{eq:p_expl} evaluated for a configuration coinciding with the real network $\mathbf{M}$ and $\{k_a\}_{a\in A}$ and $\{k_b\}_{b\in B}$ are the node degrees of $\mathbf{M}$. The best values for $\{\eta_a\}_{a\in A}$ and $\{\theta_b\}_{b\in B}$ (or equivalently $\{\mu_a\}_{a\in A}$ and $\{\nu_b\}_{b\in B}$) are therefore obtained by maximizing such log-likelihood in these parameters.
It is simple to show that this amounts to solve the system of $\|A\|+\|B\|$ equations in $\|A\|+\|B\|$  unknowns (where $\|A\|$ and $\|B\|$ simply indicate the number of nodes respectively in the two-layers $A$ and $B$ of $\mathbf{M}$):
\begin{equation}
  \left\{\begin{array}{ll}
       \sum_{b\in B} \dfrac{\eta_a\,\theta_b}{1+\eta_a\,\theta_b}=k_a \qquad\forall a\in A\\
        \sum_{a\in A} \dfrac{\eta_a\,\theta_b}{1+\eta_a\,\theta_b}=k_b\qquad\forall b\in B
        \end{array}
  \right.
  \label{max-log-like}
\end{equation}
which exactly amounts to choose $k^*_a=k_a$ $\forall a \in A$ and $k^*_b=k_b$ $\forall b \in B$.
Finally, the null model for a real bipartite network $\mathbf{M}$ is defined by Eqs.~\eqref{eq:p_expl} and \eqref{conn_prob} with the Lagrange multipliers set by Eqs.~\eqref{max-log-like}.
This recipe can therefore be applied to construct an appropriate null model for all empirical bipartite networks $\{\mathbf{M}^{\cal C, T}(y), \mathbf{M}^{\cal C, P}(y)\}_{y_{min}\le y\le y_{max}}$ obtained respectively by the PATSTAT and COMTRADE data.

In order to build a null model for the assist matrices $\mathbf{B}^{\mathcal{T\to P}}(y_1,y_2)$, and consequently for the $\Lambda$ motifs defined by Eq.~\eqref{Lambda}, we have now to compose the null models for the bipartite networks $\mathbf{M}^{\cal C, T}(y_1)$ and $\mathbf{M}^{\cal C, P}(y_2)$. This is done by contracting the two BiCMs for the matrices  $\mathbf{M}^{\cal C, T}(y_1)$ and $\mathbf{M}^{\cal C, P}(y_2)$ along the country dimension, as for Eq.~\eqref{assist}. We have:
\begin{equation}
\tilde B_{tp}(y_1,y_2)=\sum_{c=1}^{N_c}\frac{\tilde M_{ct}(y_1)}{\tilde k_t(y_1)}\frac{\tilde M_{cp}(y_2)}{\tilde k_c^{(p)}(y_2)}\,,
\label{assitz}
\end{equation}
where $\tilde k_t(y_1)=\sum_c \tilde M_{ct}(y_1)$ and $\tilde k_c^{(p)}(y_2)=\sum_p \tilde M_{cp}(y_2)$ are respectively the ubiquity of technology $t$ and the product diversification of country $c$ in the two single configurations for the BiCM null models for the two bipartite networks countries-technologies of year $y_1$ and countries-products of year $y_2$.
In other words, starting from the two BiCM ensembles for $\mathbf{M}^{\cal C, T}(y_1)$ and $\mathbf{M}^{\cal C, P}(y_2)$ we build by composition an ensemble of configurations of bipartite networks $\Omega^{\mathcal{T\to P}}(y_1,y_2)$ with link weights given by Eq.~\eqref{assitz}. 
The probability distributions of elements $\tilde B_{tp}(y_1,y_2)$, describing the null model, can be in principle obtained using exact techniques~\cite{Gualdi2016,Saracco2017}. 
However, due to the strong non-Gaussianity of such distributions, we adopt a more practical sampling technique: starting from the BiCMs for $\mathbf{M}^{\cal C, T}(y_1)$ and $\mathbf{M}^{\cal C, P}(y_2)$, we use Eqs.~\eqref{eq:p_expl}, \eqref{conn_prob} and \eqref{assitz} to generate null Assist matrices, 
and populate the related ensemble $\Omega^{\mathcal{T\to P}}(y_1,y_2)$ to estimate the full probability distributions. 
In a similar way, by using the composition Eq.~\eqref{Lambda} for the ensemble $\Omega^{\mathcal{T\to P}}(y_1,y_2)$:
\be
\tilde\Lambda_{tt'}^{p}(y_1,y_2)=\tilde B_{tp}(y_1,y_2)\tilde B_{t'p}(y_1,y_2)
\label{Lambda-null}
\ee 
and averaging over all pairs of years $y_1$ and $y_2$ with fixed delay $\Delta y=y_2-y_1$ to get $\tilde\Lambda_{tp}(\Delta y)$, we can construct the null distribution of $\Lambda$ motifs for each triple $t,t',p$ and delay $\Delta y$.

The generic observed element $\Lambda_{tt'}^{p}(\Delta y)$ is then considered statistically significant depending on the $p$-value that we can infer from its distribution under the null hypothesis. 
The specific threshold for statistical significance and the size of the generated ensemble vary on the exercises performed (as highlighted in the text). In our case we fixed the statistical significance level at $\alpha=0.01$. It is useful to recall that the two choices, the threshold and the size of the ensemble, are not unrelated: the higher the threshold we want to test, the bigger the sample we require.  We consequently extracted for each couple of years $y_1$ and $y_2$ two ensemble of $1000$ configurations/matrices for the two BiCM null models $\mathbf{\tilde M}^{\cal C, T}(y_1)$ and $\mathbf{\tilde M}^{\cal C, P}(y_2)$, and contracting one to one the configurations in the two ensembles, we get $1000$ values of $\tilde\Lambda_{tt'}^{p}(y_1,y_2)$ to finally determine the statistical significance of the observed value $\Lambda_{tt'}^{p}(\Delta y)$ as explained in Sect.~\ref{MM}.

A final comment on the issue of multiple hypothesis testing is in order here. 
Since we do many statistical tests at once (one for each motif of the Assist matrices, in order to determine the {\em true} significant elements we should use a correction for the significant threshold (such as Bonferroni or False Discovery Rate). 
However, the signal $\phi(\Delta y)$ is meant to measure the {\em average} outcome of a statistical test over all possible motifs in the network, and as such we do not need to implement any correction for the significance threshold.

\section*{Acknowledgement} This work was supported by the Italian PNR Project CRISIS-Lab. G.C. also acknowledge support from the EU project CoeGSS (676547).

\end{document}